\documentclass{article}
\usepackage{spconf,amsmath,graphicx,hyperref}
\usepackage{makecell} 
\usepackage{booktabs}  
\usepackage{multirow}  
\usepackage[table]{xcolor} 
\usepackage{enumitem}

\hypersetup{
colorlinks=true,
linkcolor=black,
citecolor=black
}

\title{Propagating Similarity, Mitigating Uncertainty: Similarity Propagation-enhanced Uncertainty for Multimodal Recommendation}
      
\name{Xinzhuo Wu$^{1,2}$ 
\qquad Hongbo Wang$^{1}$ 
\qquad Yuan Lin$^{1,\dagger}$ \thanks{$^{\dagger}$ Corresponding author. This work was supported by Sub-project of Major Consulting Research Project of Chinese Academy of Engineering (Grant No.2023-JB-10-04), National Natural Science Foundation of China (Grant No.61976036).}
\qquad Kan Xu$^{1}$
\qquad Liang Yang$^{1}$
\qquad Hongfei Lin$^{1}$
}
  
  \address{
  $^{1}$
  Dalian University of Technology,
  $^{2}$
  Zhejiang University
  }

\begin{document}
\ninept
\maketitle
\begin{abstract}
Multimodal Recommendation (MMR) systems are crucial for modern platforms but are often hampered by inherent noise and uncertainty in modal features, such as blurry images, diverse visual appearances, or ambiguous text. Existing methods often overlook this modality-specific uncertainty, leading to ineffective feature fusion. Furthermore, they fail to leverage rich similarity patterns among users and items to refine representations and their corresponding uncertainty estimates. To address these challenges, we propose a novel framework, Similarity Propagation-enhanced Uncertainty for Multimodal Recommendation (SPUMR). SPUMR explicitly models and mitigates uncertainty by first constructing the Modality Similarity Graph and the Collaborative Similarity Graph to refine representations from both content and behavioral perspectives. The Uncertainty-aware Preference Aggregation module then adaptively fuses the refined multimodal features, assigning greater weight to more reliable modalities. Extensive experiments on three benchmark datasets demonstrate that SPUMR achieves significant improvements over existing leading methods.
\end{abstract}


\section{Introduction}
Recommender systems have become indispensable for navigating the vast amount of information online, playing a crucial role in platforms like Amazon, TikTok, and YouTube. As these platforms are inherently multimodal, leveraging rich content such as images and text is essential \cite{c1}. Relying solely on user-item interaction IDs often leads to severe data sparsity and the cold-start problem, making Multimodal Recommendation (MMR) a vital research direction to enhance recommendation quality \cite{cui2025multi,cui2025diffusion}. Early MMR methods, such as VBPR \cite{c4}, incorporated multimodal data into matrix factorization but were limited by low-order interactions. Graph Neural Networks (GNNs) like MMGCN \cite{c5}, LGMRec \cite{c7}, and MIG-GT \cite{c8} addressed high-order connectivity. Self-supervised learning was introduced in BM3 \cite{c10} and LPIC \cite{c19}, but MMR systems remain vulnerable to noise. To tackle this, DA-MRS \cite{c12}, FREEDOM \cite{c13}, and BeFa \cite{c26} focus on denoising multimodal signals.

\begin{figure}[htbp]
    \centering
    \includegraphics[width=\columnwidth]{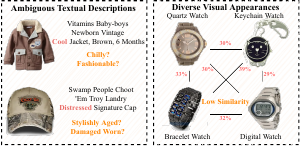}
    \vspace{-20pt}
    \caption{Illustration of the modal-specific uncertainty. The left image shows that text descriptions for products can be unclear and based on opinion, while the right shows that the same watch category can look very different (percentage represents visual similarity).}
    \label{fig:label}
    \vspace{-15pt}
\end{figure}

Despite their demonstrated success in certain scenarios, current MMR methods do not effectively mitigate the inherent uncertainty of multimodal data. Such uncertainty, originating from natural noise within each modality's feature space (e.g., blurry product images, diverse visual appearances, and ambiguous textual descriptions) \cite{lu2025dmmd4sr}, as shown in Fig.1, fundamentally undermines the performance of prevailing fusion strategies.
Although uncertainty modeling has been explored in recommendation systems \cite{c28,c29}, extending these approaches to the multimodal domain is challenging due to the following deficiencies:
\textbf{1) Difficulty in achieving uncertainty-aware multimodal fusion}: Modalities possess heterogeneous uncertainty. Some items may carry greater uncertainty in their textual modality, while others may have more in their visual modality. Yet, existing methods do not explicitly model these disparities in modal uncertainty during the fusion process, which impedes the generation of accurate and robust fused representations.
\textbf{2) Neglecting the contribution of similarity to uncertainty estimation}: In practical recommendation scenarios, similarities among users and items are highly informative—similar users tend to have similar preferences, and individual users often favor similar items. This information is crucial for constructing more precise user/item representations. Nevertheless, current methods neglect the significant role that these similarity relationships could play in improving the accuracy of uncertainty estimation.

\begin{figure*}[htbp]
    \centering
    \includegraphics[width=\textwidth]{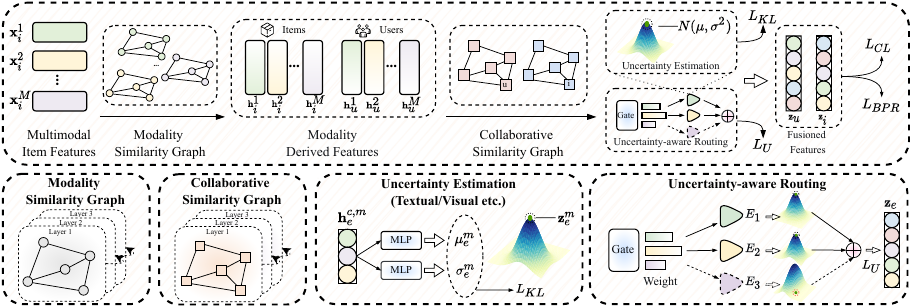}
     \vspace{-15pt}
    \caption{The overview structure of SPUMR.}
    \label{fig:label}
    \vspace{-15pt}
\end{figure*}

To address the aforementioned challenges, inspired by uncertainty modeling \cite{c23}, we propose \textbf{S}imilarity \textbf{P}ropagation-enhanced \textbf{U}ncertainty for \textbf{M}ultimodal \textbf{R}ecommendation (\textbf{SPUMR}), which effectively quantifies and models uncertainties in MMR scenarios to enhance performance. Specifically, SPUMR: 1) constructs Modality Similarity Graph and Collaborative Similarity Graph from both user and item perspectives to model multimodal and collaborative similarities, comprehensively integrating similarity information to enhance representations; 2) designs an Uncertainty-aware Preference Aggregation module that achieves precise uncertainty quantification and uncertainty-aware multimodal fusion using similarity-propagated representations.

Our main contributions are summarized as follows:

\begin{itemize}[itemsep=0pt, topsep=-2.5pt, parsep=0pt, leftmargin=1.3em]
\item We propose a novel model, SPUMR, which, to the best of our knowledge, is the first framework to incorporate uncertainty modeling into MMR, significantly enhancing its performance.
\item We design two graphs (Modality and Collaborative Similarity Graphs) to model content and collaborative similarities, which enhance representation learning while reducing noise.
\item Through extensive experimentation, we validate the effectiveness of our proposed method, and through thorough ablation studies, we verify the efficacy of each individual module.
\end{itemize}

\section{Methodology}
In this section, we describe each component in SPUMR, and the overall architecture of SPUMR is shown in Fig. 2.

\vspace{-6pt}
\subsection{Preliminary}
\label{sec:task_definition}

Let $\mathcal{U} = \{u\}$ denote the user set and $\mathcal{I} = \{i\}$ denote the item set. Then, we denote the item features of each modality as $\mathbf{E_{i}^m} \in \mathbf{R}^{|\mathcal{I}| \times d_m}$ , where $m \in \mathcal{M}$ is the modality, $\mathcal{M}$ is the set of modalities, $d_m$ is the dimension of features. In this work, we focus on the visual and textual modalities, i.e., $\mathcal{M} = \{v, t\}$, noting that our framework can be readily extended to other modalities. To handle the heterogeneity of these features, we project them into a unified latent space of dimension $d$ via a modality-specific linear transformation: $\mathbf{x}_i^{m} = \mathbf{W}^{m} \mathbf{e}_i^{m} + \mathbf{b}^{m}$, where $\mathbf{x_i}^{m} \in \mathbf{R}^{d}$.

\vspace{-6pt}
\subsection{Modality Similarity Graph}
To refine modality-specific representations and alleviate feature noise, we propagate information over graphs constructed from intra-modality similarities. This process helps to smooth feature representations by pulling similar entities closer.

\noindent
\textbf{User Modal Interest Similarity Graph.}
Users with similar tastes in a specific modality (e.g., visual aesthetics) should have aligned representations for that modality. To capture this, we first derive an initial modality-specific interest profile for each user. For a user $u$ and modality $m$, this profile $\mathbf{p}_{u}^{m}$ is computed by aggregating the features of items they have interacted with:

\vspace{-6pt}
\begin{equation}
\mathbf{p}_{u}^{m} = \frac{1}{|\mathcal{I}_u|} \sum_{i \in \mathcal{I}_u} \mathbf{x}_{i}^{m}
\end{equation}
where $\mathcal{I}_u$ is the set of items interacted with by user $u$.

We then construct a K-Nearest Neighbor (KNN) graph \cite{c24} $\mathcal{G}_{U}^{m} = (\mathcal{U}, \mathcal{E}_{U}^{m})$ for each modality $m$, where an edge exists between users $u$ and $v$ if $v$ is one of the top-k most similar users to $u$ based on the cosine similarity of their interest profiles, $\text{sim}(\mathbf{p}_{u}^{m}, \mathbf{p}_{v}^{m})$.We perform graph propagation on $\mathcal{G}_{U}^{m}$ to refine user modal representations. Let $\mathbf{h}_{u}^{(m, 0)} = \mathbf{p}_{u}^{m}$ be the initial representation. The propagation at layer $l$ is defined as:  

\vspace{-10pt}
\begin{equation}
\mathbf{h}_{u}^{(m, l+1)} = \sum_{v \in \mathcal{N}_{u}^{m} \cup \{u\}} \frac{1}{\sqrt{|\mathcal{N}_{u}^{m}|+1} \sqrt{|\mathcal{N}_{v}^{m}|+1}} \mathbf{h}_{v}^{(m, l)}
\end{equation}
where $\mathcal{N}_{u}^{m}$ is the set of neighbors of user $u$ in $\mathcal{G}_{U}^{m}$. After $L$ layers, we obtain the refined user representation for modality $m$, denoted as $\mathbf{h}_{u}^{m} = \mathbf{h}_{u}^{(m, L)}$.

\noindent
\textbf{Item Modal Semantic Similarity Graph.}
Similarly, items that are semantically related in a specific modality (e.g., two visually similar red shirts) should have similar representations. For each modality $m$, we construct an item-item KNN graph $\mathcal{G}_{I}^{m} = (\mathcal{I}, \mathcal{E}_{I}^{m})$ based on the cosine similarity of their initial modal features $\{\mathbf{x}_{i}^{m}\}_{i \in \mathcal{I}}$.

Analogous to user graph propagation, we refine the item modal representations by propagating information over $\mathcal{G}_{I}^{m}$. Starting with $\mathbf{h}_{i}^{(m, 0)} = \mathbf{x}_{i}^{m}$, the refined item representation for modality $m$ is obtained as:

\vspace{-10pt}
\begin{equation}
\mathbf{h}_{i}^{(m, l+1)} = \sum_{j \in \mathcal{N}_{i}^{m} \cup \{i\}} \frac{1}{\sqrt{|\mathcal{N}_{i}^{m}|+1} \sqrt{|\mathcal{N}_{j}^{m}|+1}} \mathbf{h}_{j}^{(m, l)}
\end{equation}

The final representation after $L$ layers is denoted as $\mathbf{h}_{i}^{m} = \mathbf{h}_{i}^{(m, L)}$.

\vspace{-6pt}
\subsection{Collaborative Similarity Graph}
To complement modal semantics, we explicitly model the collaborative filtering signal derived from user-item interactions. This captures behavioral patterns that are often orthogonal to content features. Crucially, instead of collapsing modal information prematurely, we use the collaborative structure to refine each modal representation space independently.

\noindent
\textbf{User Collaborative Similarity Graph.}
Users who interact with a similar set of items are likely to share common preferences. We construct a user-user collaborative similarity graph $\mathcal{G}_{U}^{c} = (\mathcal{U}, \mathcal{E}_{U}^{c})$. The similarity between two users $u$ and $v$ is calculated using the Jaccard similarity of their interaction sets: $w_{uv} = \frac{|\mathcal{I}_u \cap \mathcal{I}_v|}{|\mathcal{I}_u \cup \mathcal{I}_v|}$. Edges are formed between a user and their top-k most similar peers. We then leverage this collaborative graph to enhance each of the user's modal representations. For each modality $m \in \mathcal{M}$, we perform a separate graph propagation on $\mathcal{G}_{U}^{c}$, initializing the node features with the representations $\mathbf{h}_{u}^{m}$ obtained from the modal-semantic propagation step. The one-layer propagation is defined as:

\vspace{-6pt}
\begin{equation}
\mathbf{h}_{u}^{c,m} = \sum_{v \in \mathcal{N}_{u}^{c} \cup \{u\}} \tilde{w}_{uv} \mathbf{h}_{v}^{m}
\end{equation}
where $\mathcal{N}_{u}^{c}$ is the neighborhood of user $u$ in $\mathcal{G}_{U}^{c}$, and $\tilde{w}_{uv}$ is the normalized similarity score, e.g., $\tilde{w}_{uv} = w_{uv} / \sum_{k \in \mathcal{N}_{u}^{c} \cup \{u\}} w_{uk}$. This process yields a set of collaboratively aware user representations, denoted as $\{\mathbf{h}_{u}^{c,m}\}_{m \in \mathcal{M}}$.

\noindent
\textbf{Item Collaborative Similarity Graph.}
Items frequently co-interacted with by the same users often exhibit semantic or functional relationships (e.g., shirts and pants). We build an item-item collaborative graph $\mathcal{G}_{I}^{c} = (\mathcal{I}, \mathcal{E}_{I}^{c})$ based on item co-occurrence. The similarity between items $i$ and $j$ is defined as $w_{ij} = \frac{|\mathcal{U}_i \cap \mathcal{U}_j|}{|\mathcal{U}_i \cup \mathcal{U}_j|}$, where $\mathcal{U}_i$ is the set of users who interacted with item $i$. Mirroring the process for users, we refine each item's modal representations using the item collaborative graph $\mathcal{G}_{I}^{c}$. For each modality $m \in \mathcal{M}$, we initialize the propagation with the features $\mathbf{h}_{i}^{m}$ and diffuse them over the graph:

\vspace{-10pt}
\begin{equation}
\mathbf{h}_{i}^{c,m} = \sum_{j \in \mathcal{N}_{i}^{c} \cup \{i\}} \tilde{w}_{ij} \mathbf{h}_{j}^{m}
\end{equation}

This results in a set of collaboratively aware item representations $\{\mathbf{h}_{i}^{c,m}\}_{m \in \mathcal{M}}$, which now embed co-interaction patterns into each distinct modality.

\vspace{-6pt}
\subsection{Uncertainty-aware Preference Aggregation}
\label{sec:uncertainty_aggregation}

The refined modality-specific representations $\{ \mathbf{h}_{e}^{c,m} \}_{m \in \mathcal{M}}$ for each entity $e \in \mathcal{U} \cup \mathcal{I}$ may carry varying degrees of noise. To address this, we introduce an uncertainty-based fusion mechanism that adaptively aggregates these representations based on their modeled uncertainty.

\noindent
\textbf{Uncertainty Estimation.}
We model the uncertainty of each representation $\mathbf{h}_{e}^{c,m}$ by treating it as a sample from a latent Gaussian distribution \cite{c25}. Two MLPs, $\text{MLP}_{\mu}$ and $\text{MLP}_{\sigma}$, parameterize the mean $\mu_e^m$ and variance $(\sigma_e^m)^2$:

\vspace{-12pt}
\begin{equation}
\mu_e^m = \text{MLP}_{\mu}(\mathbf{h}_{e}^{c,m}), \quad \log(\sigma_e^m)^2 = \text{MLP}_{\sigma}(\mathbf{h}_{e}^{c,m})
\end{equation}

We apply the reparameterization trick to obtain differentiable stochastic embeddings $\mathbf{z}_e^m = \mu_e^m + \sigma_e^m \odot \epsilon$ for differentiable sampling, where $\epsilon \sim \mathcal{N}(0, \mathbf{I})$. To prevent variance collapse, we apply a Kullback-Leibler (KL) divergence loss, $\mathcal{L}_{KL}$, between the learned distribution and a standard normal prior.

\vspace{-6pt}
\begin{equation}
\mathcal{L}_{KL} = \sum_{e, m} \text{KL}(\mathcal{N}(\mu_e^m, \text{diag}((\sigma_e^m)^2)) || \mathcal{N}(0, \mathbf{I}))
\end{equation}

\noindent
\textbf{Uncertainty-aware Routing.}
We aggregate the stochastic modal embeddings ("experts") using an uncertainty-aware routing mechanism. A gating network $G$ computes modality-specific weights, assigning lower importance to experts with higher uncertainty. The gate takes the concatenated representations $\{\mathbf{h}_e^{c,m}\}_{m \in \mathcal{M}}$ as input and employs a Top-K function to encourage specialization:

\vspace{-12pt}
\begin{equation}
G(\mathbf{h}_e^{c}) = \text{softmax}(\text{Top-K}(\mathbf{W}_g [\mathbf{h}_{e}^{c,m_1}; \dots; \mathbf{h}_{e}^{c,m_{|\mathcal{M}|}}]))
\end{equation}

The final aggregated representation $\mathbf{z}_e$ is a weighted sum of the expert outputs:
$
\mathbf{z}_e = \sum_{m \in \mathcal{M}} G(\mathbf{h}_e^{c})_m \cdot \mathbf{z}_e^m
$.
To explicitly guide the gate, we introduce an uncertainty regularization loss, $\mathcal{L}_{U}$, which penalizes assigning high weights to high-variance experts:

\vspace{-6pt}
\begin{equation}
\mathcal{L}_{U} = \sum_{e \in \mathcal{U} \cup \mathcal{I}} \sum_{m \in \mathcal{M}} G(\mathbf{h}_e^{c})_m \cdot ||\sigma_e^m||_2^2
\end{equation}

\noindent
\textbf{Prediction and Optimization.}
The final user $\mathbf{z}_u$ and item $\mathbf{z}_i$ representations are learned by optimizing a composite loss function, which includes four components. First, a Bayesian Personalized Ranking (BPR) loss is used for optimizing the ranking objective:

\vspace{-10pt}
\begin{equation}
\mathcal{L}_\text{BPR} = \sum_{u, i} \log ( \sigma(\mathbf{z}_u \cdot \mathbf{z}_i) ) - \sum_{u, i'} \log ( \sigma(-\mathbf{z}_u \cdot \mathbf{z}_{i'}) )
\end{equation}
\vspace{-12pt}

\noindent
where $i$ denotes the positive item and $i'$ denotes the negative item sampled for user $u$. Second, a contrastive loss pulls the representations of a user and their positive items closer:

\vspace{-5pt}
\begin{equation}
\mathcal{L}_\text{CL} = -\log \left( \frac{\exp(\mathbf{z}_u \cdot \mathbf{z}_i)}{ \sum_{i' \in \mathcal{I}_u} \exp(\mathbf{z}_u \cdot \mathbf{z}_{i'})} \right)
\end{equation}

Finally, the overall objective function combines these losses with our proposed uncertainty regularizers ($\mathcal{L}_{KL}$ and $\mathcal{L}_{U}$):

\vspace{-5pt}
\begin{equation}
\mathcal{L} = \mathcal{L}_\text{BPR} + \lambda_\text{CL} \mathcal{L}_\text{CL} + \lambda_\text{KL} \mathcal{L}_\text{KL} + \lambda_U \mathcal{L}_U
\end{equation}
where $\lambda_{CL}$, $\lambda_{KL}$, and $\lambda_{U}$ are balancing hyperparameters.

\section{Experiments}
To validate the effectiveness of SPUMR, we propose the following four research questions to guide our experiments: \textbf{RQ1}: How effective is SPUMR compared to the state-of-the-art multimodal recommender systems? \textbf{RQ2}: How do the core components of SPUMR affect the model performance? \textbf{RQ3}: How do the hyper-parameter variations impact the performance of SPUMR? \textbf{RQ4}: How does the uncertainty mechanism improve the model's robustness to noisy modal features?

\vspace{-6pt}
\subsection{Experiment Setup}
\noindent
\textbf{Datasets and Evaluation Metrics.} Following prior
work \cite{c5}, we conduct experiments on three public benchmark datasets from 5-core Amazon: Baby, Sports, and Clothing \cite{c16}. The detailed statistics of the datasets are summarized in Table 1. For each user, the interaction data is split into training, validation, and test sets in an 8:1:1 ratio. We evaluate model performance using two standard Top-K ranking metrics: Recall@K (R@K) and NDCG@K (N@K). These metrics assess the presence and position of ground-truth items within the Top-K recommendation list. We report the average results across all test users for K = {10, 20}.

\vspace{-15pt}
\begin{table}[htbp]
\centering
\caption{Statistics of datasets.}
\label{tab:dataset_stats}
\begin{tabular}{ccccc}
\toprule
\textbf{Datasets} & \textbf{\#Users} & \textbf{\#Items} & \textbf{\#Interactions} & \textbf{Sparsity} \\
\midrule
\textbf{Baby}     & 19,445           & 7,050            & 160,792                 & 99.88\%           \\
\textbf{Sports}   & 35,598           & 18,357           & 296,337                 & 99.95\%           \\
\textbf{Clothing}   & 39,387            & 23,033            & 278,677                  & 99.97\%           \\
\bottomrule
\end{tabular}
\vspace{-8pt}
\end{table}

\begin{table*}[htbp]
  \centering 
  \caption{Performance comparison of baselines and SPUMR(ours) in terms of Recall@K(R@K) and NDCG@K(N@K). The superscript * indicates the improvement is statistically significant where the p-value is less than 0.01.}
  \definecolor{lightgray}{gray}{0.9}
  \setlength{\tabcolsep}{2.5pt}
  \renewcommand{\arraystretch}{0.96} 
  \begin{tabular}{rcccccccccccc}
    \toprule
    \multirow{2}{*}{\textbf{Model (Source)}} & \multicolumn{4}{c}{\textbf{Baby}} & \multicolumn{4}{c}{\textbf{Sports}} & \multicolumn{4}{c}{\textbf{Clothing}} \\
    \cmidrule(lr){2-5} \cmidrule(lr){6-9} \cmidrule(lr){10-13}
    & R@10 & R@20 & N@10 & N@20 & R@10 & R@20 & N@10 & N@20 & R@10 & R@20 & N@10 & N@20 \\
    \hline
    MF-BPR (UAI'09) & 0.0357 & 0.0575 & 0.0192 & 0.0249 & 0.0432 & 0.0653 & 0.0241 & 0.0298 & 0.0187 & 0.0279 & 0.0103 & 0.0126 \\
    LightGCN (SIGIR'20) & 0.0479 & 0.0754 & 0.0257 & 0.0328 & 0.0569 & 0.0864 & 0.0311 & 0.0387 & 0.0340 & 0.0526 & 0.0188 & 0.0236 \\
    \hline
    OrdRec (TORS'23) & 0.0509 & 0.0798 & 0.0266 & 0.0347 & 0.0596 & 0.0911 & 0.0324 & 0.0410 & 0.0351 & 0.0545 & 0.0192 & 0.0241
    \\
    VAE-AUR (TBD'23) & 0.0528 & 0.0834 & 0.0287 & 0.0369 & 0.0618 & 0.0934 & 0.0331 & 0.0421 & 0.0376 & 0.0572 & 0.0211 & 0.0260
    \\
    \hline
    VBPR (AAAI'16) & 0.0423 & 0.0663 & 0.0223 & 0.0284 & 0.0558 & 0.0856 & 0.0307 & 0.0384 & 0.0281 & 0.0415 & 0.0158 & 0.0192 \\
    MMGCN (MM'19) & 0.0378 & 0.0615 & 0.0200 & 0.0261 & 0.0370 & 0.0605 & 0.0193 & 0.0254 & 0.0218 & 0.0345 & 0.0110 & 0.0142 \\
    FREEDOM (MM'23) & 0.0627 & 0.0992 & 0.0330 & 0.0424 & 0.0717 & 0.1089 & 0.0385 & 0.0481 & 0.0628 & 0.0941 & 0.0341 & 0.0420 \\
    BM3 (WWW'23) & 0.0564 & 0.0883 & 0.0301 & 0.0383 & 0.0656 & 0.0980 & 0.0355 & 0.0438 & 0.0422 & 0.0621 & 0.0231 & 0.0281 \\
    LGMRec (AAAI'24) & 0.0644 & 0.1002 & 0.0349 & 0.0440 & 0.0720 & 0.1068 & 0.0390 & 0.0480 & 0.0555 & 0.0828 & 0.0302 & 0.0371 \\
    DA-MRS (KDD'24) & 0.0650 & 0.0994 & 0.0346 & 0.0435 & 0.0751 & 0.1125 & 0.0402 & 0.0498 & \underline{0.0647} & \underline{0.0963} & \underline{0.0353} & \underline{0.0433} \\
    BeFA (AAAI'25) & 0.0555 & 0.0884 & 0.0299 & 0.0383 & 0.0649 & 0.0985 &0.0346 &0.0432 & 0.0568 & 0.0857 & 0.0307 & 0.0381
    \\
    LPIC (TOMM'25) & 0.0634 & 0.0977 & 0.0337 & 0.0422 & 0.0737 & 0.1113 &0.0398 &0.0485 &0.0627 & 0.0928 & 0.0338 & 0.0405
    \\
    MIG-GT (AAAI'25) & \underline{0.0665} & \underline{0.1021} & \underline{0.0361} & \underline{0.0452} & \underline{0.0753} & \underline{0.1130} & \underline{0.0414} & \underline{0.0511} & {0.0636} & {0.0934} & {0.0347} & {0.0422}
    \\
    \hline
    \textbf{SPUMR (Ours
    )} & \textbf{0.0711}* & \textbf{0.1073}* & \textbf{0.0385}* & \textbf{0.0477}* & \textbf{0.0788}* & \textbf{0.1168}* & \textbf{0.0442}* & \textbf{0.0539}* & \textbf{0.0688}* & \textbf{0.1004}* & \textbf{0.0379}* & \textbf{0.0458}* \\
    \rowcolor{lightgray}
    \textit{Improv.} & \textit{6.92\%} & \textit{5.09\%} & \textit{6.65\%} & \textit{5.53\%} & \textit{4.65\%} & \textit{3.36\%} & \textit{6.76\%} & \textit{5.48\%} & \textit{6.34\%} & \textit{4.26\%} & \textit{7.37\%} & \textit{5.77\%} \\
    \bottomrule
  \end{tabular}%
\vspace{-11pt}
\end{table*}

\noindent
\textbf{Baselines and Implementation Details.} To evaluate the effectiveness of SPUMR , we compare it with state-of-the-art recommendation models, categorized into three groups: conventional methods (MF-BPR \cite{c2} and LightGCN \cite{c15}), uncertainty-aware models (OrdRec \cite{c28} and VAE-AUR \cite{c29}), and multimodal approaches (VBPR \cite{c4}, MMGCN \cite{c5},  FREEDOM \cite{c13}, BM3 \cite{c10}, LGMRec \cite{c7}, DA-MRS \cite{c12}, BeFA \cite{c26}, LPIC \cite{c19}, and MIG-GT \cite{c8}).

Our model, along with all baselines, is trained within the MMRec framework to ensure a fair comparison \cite{c17}. We use a batch size of 2048, a learning rate of 0.001, Xavier initialization, and the Adam optimizer. Following prior work \cite{c18}, we set the embedding dimension to $d=64$. For the similarity graphs (Modality and Collaborative), we use a depth of $L=3$ and $k=10$ neighbors. For Uncertainty-aware Preference Aggregation, $\lambda_{KL}$ and $\lambda_{U}$ are grid-searched in $\{1\text{e-}2, 5\text{e-}3, 1\text{e-}3, 5\text{e-}4, 1\text{e-}4\}$, while $\lambda_{CL}$ is fixed at 0.01. Training employs early stopping at 20 epochs based on Recall@20. All experiments are implemented in PyTorch and conducted on a single NVIDIA A100 GPU with 80GB of memory.

\vspace{-6pt}
\subsection{Overall Performance (RQ1)}
Detailed experiment results are shown in Table 2. The optimal results are highlighted in bold, while the suboptimal ones are underlined. More specifically, based on these results,  we observed that SPUMR outperforms the strongest baselines, achieving 6.65\%(N@10), 5.53\%(N@20) improvement on the Baby dataset, 6.76\%(N@10), 5.48\%(N@20) improvement on the Sports dataset, and 7.37\%(N@10), 5.77\%(N@20) improvement on the Clothing dataset, which demonstrates the effectiveness of SPUMR.

\vspace{-8pt}
\subsection{Ablation Study (RQ2)}
We conduct an ablation study to evaluate each key component of SPUMR by individually removing Modality Similarity Graph (w/o MSG), Collaborative Similarity Graph (w/o CSG), and Uncertainty-aware Preference Aggregation (w/o UPA). The results, presented in Table 3, demonstrate that removing any of these three components leads to a degradation in performance across all datasets and metrics.

\vspace{-15pt}
\begin{table}[htbp]
  \centering
  \caption{Ablation study of SPUMR. Performance is evaluated on three datasets using R@20 and N@20.}
  \definecolor{lightgray}{gray}{0.9}
  \setlength{\tabcolsep}{3.8pt}
  \renewcommand{\arraystretch}{0.75} 
  \begin{tabular}{ccccccc}
    \toprule
    \multirow{2}{*}{\textbf{\makecell{Model \\ Variant}}} & \multicolumn{2}{c}{\textbf{Baby}} & \multicolumn{2}{c}{\textbf{Sports}} & \multicolumn{2}{c}{\textbf{Clothing}} \\
    \cmidrule(lr){2-3} \cmidrule(lr){4-5} \cmidrule(lr){6-7}
    & R@20 & N@20 & R@20 & N@20 & R@20 & N@20 \\
    \midrule
    \rowcolor{lightgray}
    SPUMR & \textbf{0.1073} & \textbf{0.0477} & \textbf{0.1168} & \textbf{0.0539} & \textbf{0.1004} & \textbf{0.0458} \\
    w/o MSG & 0.1055 & 0.0459 & 0.1149 & 0.0523 & 0.0983 & 0.0444 \\
    w/o CSG & 0.1053 & 0.0456 & 0.1146 & 0.0519 & 0.0978 & 0.0440 \\
    w/o UPA & 0.1066 & 0.0469 & 0.1158 & 0.0527 & 0.0989 & 0.0450 \\
    \bottomrule
  \end{tabular}
\vspace{-18pt}
\end{table}

\begin{figure}[htbp]
    \centering
    \includegraphics[width=\columnwidth]{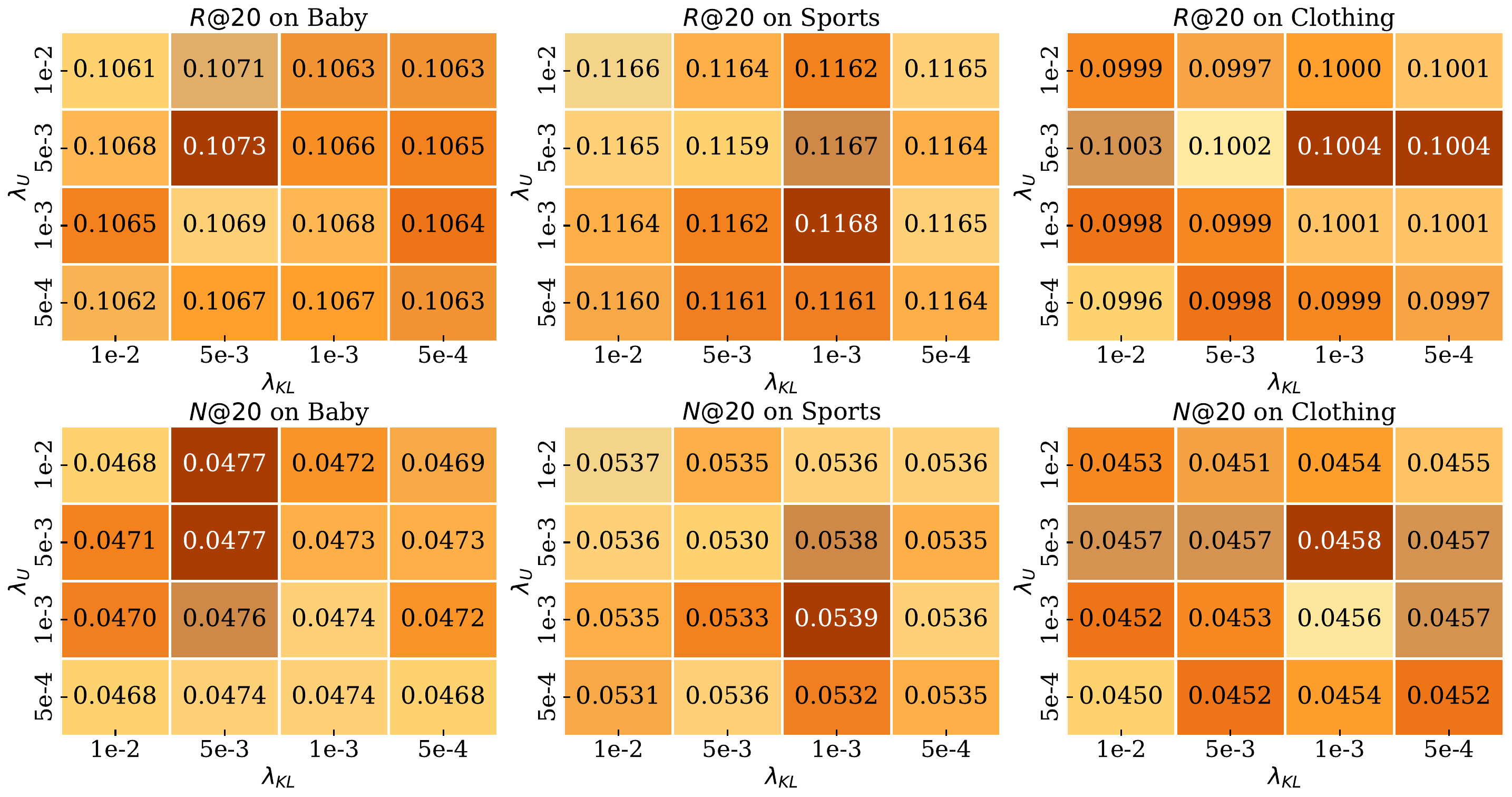}
    \vspace{-20pt}
    \caption{Performance of SPUMR under different values of $\lambda_{KL}$ and $\lambda_{U}$ on three datasets.}
    \label{fig:label}
    \vspace{-10pt}
\end{figure}

\vspace{-6pt}
\subsection{Hyper-parameter Analysis (RQ3)}
To examine the sensitivity of SPUMR to hyper-parameters, we evaluated its performance on three datasets with different hyper-parameter values. Fig. 3 indicates that: for $\lambda_{KL}$, the optimal value is 5e-3 on the Baby dataset and 1e-3 on Sports and Clothing;
for $\lambda_{U}$, 5e-3 performs best on Baby and Clothing, while 1e-3 is optimal for Sports. Notably, flexible hyper-parameter selection enhances model adaptability across diverse datasets.

\vspace{-6pt}
\subsection{Visualization (RQ4)}
We randomly sample 500 users from the Baby dataset and apply t-SNE to visualize the learned user representations in 2D space \cite{c20}. As shown in Fig. 4, the embeddings learned by MMGCN or FREEDOM exhibit localized clustering, while SPUMR achieves a more uniform distribution. Previous research \cite{c27} suggests that such decentralized representations better preserve key features, improving recommendation accuracy and robustness.

\vspace{-6.5pt}
\begin{figure}[htbp]
    \centering
    \includegraphics[width=\columnwidth]{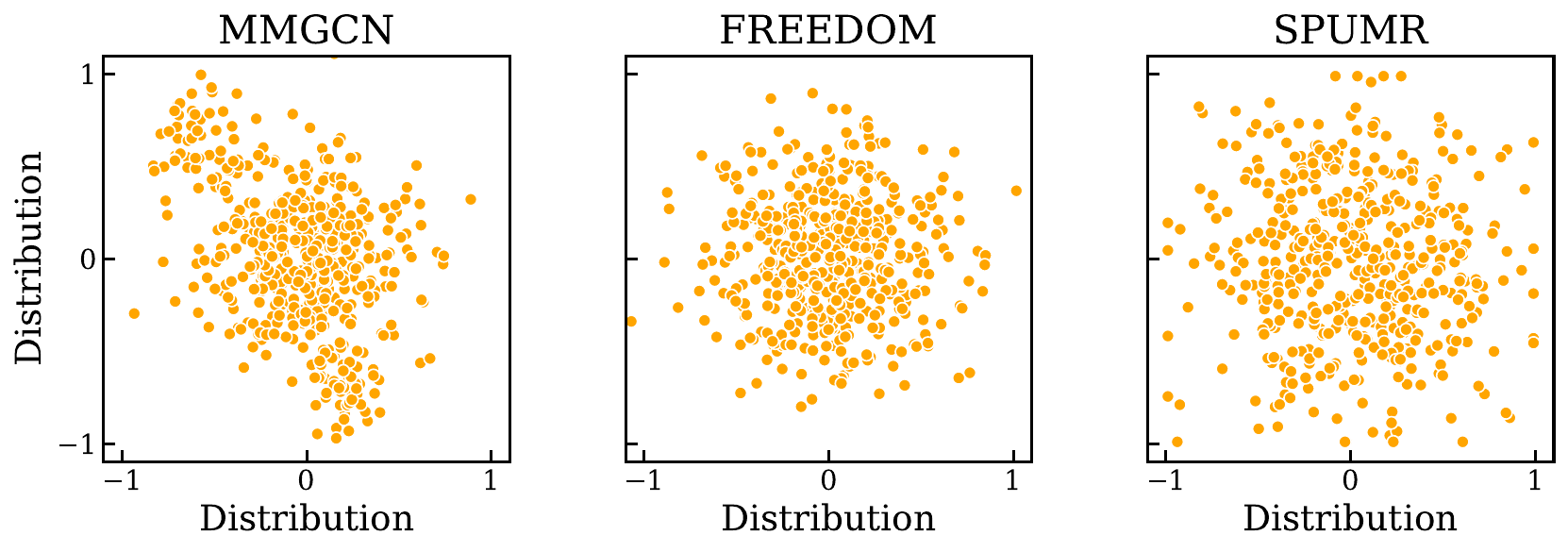}
     \vspace{-22pt}
    \caption{The distribution of the user representations learned by MMGCN, FREEDOM, and SPUMR on the Baby dataset.}
    \label{fig:label}
    \vspace{-10pt}
\end{figure}

\section{CONCLUSION}
This paper introduces SPUMR, a novel framework for stochastic uncertainty in Multimodal Recommendation. SPUMR denoises representations by propagating information across modality and collaborative similarity graphs, and then adaptively fuses them based on uncertainty to mitigate noise. Extensive experiments on three benchmark datasets show that SPUMR significantly outperforms leading methods. Future work could explore integrating our framework with multimodal large models.

\vfill\pagebreak
\bibliographystyle{IEEEbib}
{
\small
\bibliography{strings,refs}
}
\end{document}